\documentclass[lettersize,journal]{IEEEtran}
\usepackage{amsmath,amsfonts,amssymb}
\usepackage{array}
\usepackage{textcomp}
\usepackage{stfloats}
\usepackage{url}
\usepackage{graphicx}
\usepackage{cite}
\usepackage{booktabs}
\usepackage[dvipsnames]{xcolor}
\usepackage{amsthm}
\newcommand{\SINR}{\text{SINR}}
\newcommand{\figref}[1]{Fig.~\ref{#1}}
\newcommand{\tabref}[1]{Table~\ref{#1}}
\newcommand{\secref}[1]{Section~\ref{#1}}
\def\cB{\mathcal{B}}
\def\cC{\mathcal{C}}
\def\cG{\mathcal{G}}
\def\cT{\mathcal{T}}
\newcommand{\setB}{\cB}
\newcommand{\setT}{\cT}
\newcommand{\setG}{\cG}
\newcommand{\actsym}{x}
\newcommand{\ptx}{p^{\text{tx}}}
\newcommand{\pch}{p^{\text{ch}}}
\newcommand{\pdis}{p^{\text{dis}}}
\newcommand{\pgrid}{P^{\text{grid}}}
\newcommand{\xv}{\actsym_{t,b,g}}
\newcommand{\ptxv}{\ptx_{t,b,g}}
\newcommand{\Bv}{B_{t,b,g}}
\newcommand{\Bvl}{B_{t,b(\ell),g(\ell)}}
\newcommand{\ptxvl}{\ptx_{t,b(\ell),g(\ell)}}
\newcommand{\Wtot}{W^{\text{tot}}}
\newcommand{\PmaxBS}{P^{\text{tx}}_{\max}}
\newcommand{\SINRtl}{\SINR_{t,\ell}}
\newcommand{\Rtl}{R_{t,\ell}}
\newcommand{\Dtl}{D_{t,\ell}}
\newcommand{\Pcomm}{P^{\text{comm}}_{t,b}}
\newcommand{\Psta}{P^{\text{sta}}_{b}}
\newcommand{\Pact}{P^{\text{act}}_{b}}
\newcommand{\pgridv}{\pgrid_{t,b}}
\newcommand{\pchv}{\pch_{t,b}}
\newcommand{\pdisv}{\pdis_{t,b}}
\newcommand{\Ev}{E_{t,b}}
\newcommand{\Emax}{E_{\max}}
\newcommand{\Pchmax}{P^{\text{ch}}_{\max}}
\newcommand{\Pdismax}{P^{\text{dis}}_{\max}}
\newcommand{\etac}{\eta_{\text{c}}}
\newcommand{\etad}{\eta_{\text{d}}}
\newcommand{\piEff}{\tilde{\pi}_t}

\newcommand{\caisoDate}{Apr.~15, 2026}
\newcommand{\caisoNode}{SP15}
\newcommand{\priceRange}{$-10.6$ to $25.3$~USD/MWh}
\newcommand{\ciRange}{62 to 213~gCO$_2$/kWh}
\newcommand{\cpNow}{28.76}
\newcommand{\ablEmisCutNow}{7.3\%}
\newcommand{\ablEmisCutRange}{7\%}
\newcommand{\cpKneeLow}{100}
\newcommand{\cpKneeHigh}{300}
\newcommand{\emisAtKnee}{1.16}
\newcommand{\slotGapMax}{2\%}
\newcommand{\qosMargin}{0.05}
\newcommand{\robWinterCut}{7.7\%}
\newcommand{\caisoDateAug}{Aug.~15, 2026}
\newcommand{\ablEmisCutAug}{37\%}
\newcommand{\augBelowNoBattery}{59\%}
\newcommand{\augElecPremium}{3\%}
\newcommand{\robAugCut}{57.8\%}

\begin{document}
\bstctlcite{IEEEexample:BSTcontrol}

\title{Carbon-Aware Joint Resource and Energy Scheduling for Cellular Base Stations}

\author{Jaebum Park,~\IEEEmembership{Graduate Student Member,~IEEE}, Yutao Xie, Amitava Ghosh,~\IEEEmembership{Fellow,~IEEE}\\
  Yuanyuan Shi,~\IEEEmembership{Member,~IEEE},
  and Robert W. Heath Jr.,~\IEEEmembership{Fellow,~IEEE}%

\thanks{This work was supported in part by a gift from the Nokia Foundation.}
\thanks{J. Park, Y. Xie, Y. Shi, and R. W. Heath, Jr.\ are with the Department of
Electrical and Computer Engineering, University of California at San Diego,
La Jolla, CA 92093, USA (emails: \{jp261, yux076, yyshi, rwheathjr\}@ucsd.edu).
}
\thanks{A. Ghosh is with Nokia Standards, Naperville, IL 60563, USA (email: amitava.ghosh@nokia.com).}
}

\maketitle

\begin{abstract}
Decarbonizing cellular networks requires coordinating communication-resource and battery operation with time-varying electricity prices and carbon intensity. We jointly optimize base station and battery operation under a carbon-adjusted electricity price and traffic-demand constraints, and solve the mixed-integer nonconvex problem through a two-stage decomposition exact under a minimum-power rule. Using California price and carbon-intensity data, we show that carbon-aware scheduling reduces emissions most when low-price and low-carbon periods are misaligned. Above a carbon price of roughly \cpKneeLow--\cpKneeHigh~USD/ton, battery operation shifts from cost-driven to carbon-driven, cutting emissions by up to \robAugCut{} relative to a cost-only schedule.
\end{abstract}


\section{Introduction}

\IEEEPARstart{B}{ase} stations account for about 73\% of a mobile network's energy consumption~\cite{GreenFutureNetworks2021}, with the radio unit taking a growing share as massive multiple-input multiple-output (MIMO) scales up~\cite{WesemannEtAlEnergyEfficientExtreme2023}. Base station operation is therefore an actionable lever for reducing both operational expenditure and emissions, and energy-efficiency indicators now appear throughout 5G-Advanced and 6G roadmaps~\cite{3gppTR38.864,3gpp:6GWorkshopReport2025}. Traffic migrates across cells with commuting patterns~\cite{wang2018spatio}, while electricity prices and grid carbon intensity vary over the day. The radio resources follow the traffic in each slot, and the battery shifts grid purchases toward the cheap and clean hours, so the same demand is served at lower cost and lower emissions. 

Prior work couples the communications and energy sides only in part. On the wireless communications side, sleep control, carrier muting, and antenna switching reduce consumption under time-varying load~\cite{3gppTR38.864}, and our prior cross-layer design jointly optimizes the rate adaptation and power consumption of virtualized RAN components~\cite{ParkEtAlEnergyEfficiencyOptimization2026}. This line does not model where the energy comes from or what it costs. Smart-grid studies schedule grid import and battery operation for base stations under time-varying prices~\cite{BuEtAlWhenSmartGrid2012}, but abstract the communications side into a fixed load. The spatial--temporal energy management framework in~\cite{ZhouEtAlSPAtialTemporalEnergy2022} couples association with energy scheduling, but optimizes electricity cost alone. Carbon-aware formulations add emission signals to network control~\cite{YangEtAlCarbonneutralizedJointUser2023,cheng2022carbon,ChenEtAlCarbonEmissions2024}, but do not quantify the carbon price at which carbon awareness changes what a base station does. For a battery alone, a carbon price is known to align arbitrage with emissions~\cite{OlsenKirschen2020}, but with the load taken as given, whereas the load of a base station follows the traffic.

In this letter, we formulate day-ahead operation of cellular base stations as a single monetized program that internalizes carbon emissions through a carbon-adjusted electricity price. We couple subband activation, bandwidth and power allocation, and battery scheduling under per zone rate constraints. The communication plan follows the demand and does not depend on the price. It sets the load that the battery shifts across the day. The demand comes from the agent-based mobility pipeline of our prior work~\cite{ParkEtAlEnergyEfficiencyOptimization2026}, which extends MATSim~\cite{w2016multi}. We decompose the resulting mixed-integer nonconvex program into per slot resource subproblems and a battery mixed-integer linear program. The decomposition handles a full day at 15-minute resolution, with each subproblem solved to an optimality gap of at most \slotGapMax{}. We then compare with a cost-only schedule across storage capacities and sweep the carbon price to locate the threshold at which the battery schedule responds.

With California price and carbon-intensity data, current carbon prices reduce daily emissions by only about \ablEmisCutRange{} relative to a cost-only schedule when the price and carbon valleys coincide, as on a spring day with negative midday prices. When the valleys separate, as on a summer day, the same prices cut emissions by \ablEmisCutAug{} for a \augElecPremium{} increase in electricity cost. Once the carbon price exceeds roughly \cpKneeLow--\cpKneeHigh~USD per ton, charging aligns with carbon on every day we test, and the attainable cut spans \robWinterCut{} in winter to \robAugCut{} in summer.\looseness=-1

\section{System model}
\label{sec:system}

\begin{figure}[t]
  \centering
  \includegraphics[width=0.88\columnwidth]{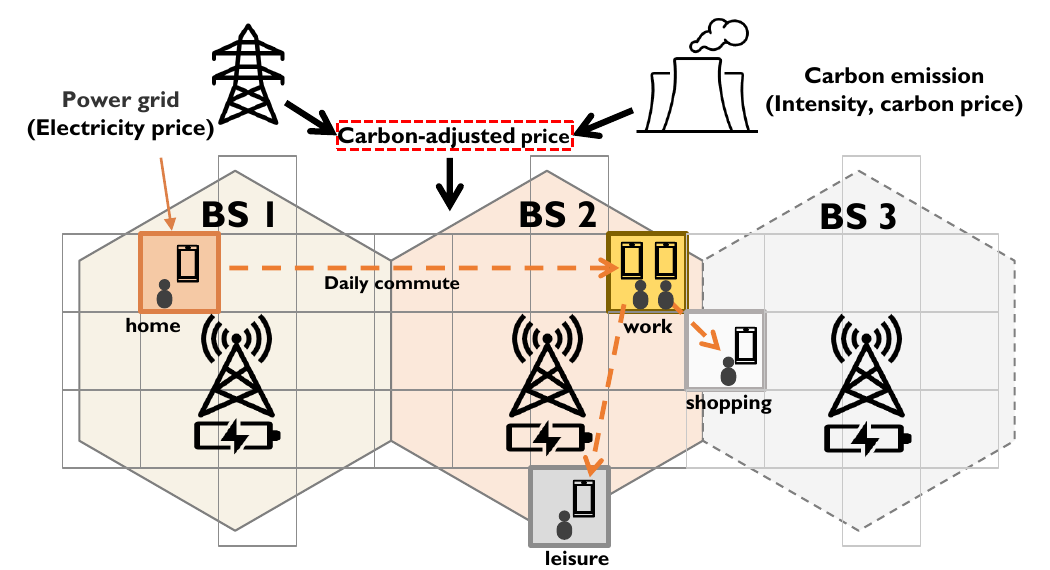}
  \caption{System model. Agents moving between home, work, shopping, and leisure cells create the zone demand, which each base station serves while drawing power from the grid and a local battery. The carbon surcharge is added to the grid electricity price to form the carbon-adjusted price.}
  \label{fig:system-model}
\end{figure}

We consider a network of base stations $b\in\setB$ in a two-dimensional hexagonal urban-micro layout. In each discrete time slot $t\in\setT$ of duration $\tau$ hours, a controller selects the active subbands, their bandwidth and transmit power, and the battery action at every base station, to minimize the cost of electricity and monetized carbon. \figref{fig:system-model} sketches the architecture.

\subsection{Demand and serving map}
\label{sec:mobility-demand}

Base stations sit at the centers of the hexagonal cells. To model demand, we discretize the covered area into $L$ square zones indexed by $\ell$, and time into $T$ slots.
An agent-based mobility pipeline~\cite{ParkEtAlEnergyEfficiencyOptimization2026}, which extends the multi-agent transport simulator~\cite{w2016multi}, moves the users between daily activity locations of four types, that is, home, work, shopping, and leisure, which we anchor in designated hexagonal cells. Let $N_{t,\ell}$ denote the time-averaged number of users in zone $\ell$ during slot $t$. Let $\cC$ denote a set of three empirical traffic classes, indexed by $c$, each defined by the access point name (APN) under which its traffic was recorded in the operator trace of~\cite{ParkEtAlEnergyEfficiencyOptimization2026} and each with a diurnal per user demand profile $u_c(t)$. The weight $w_c(\ell)\ge 0$ is the fraction of the traffic in zone $\ell$ that belongs to class $c$, set by the activity type anchored at that location. The weights sum to one over $\cC$, so zone $\ell$ offers the demand
\begin{equation}
D_{t,\ell} = N_{t,\ell}\sum_{c\in\cC} w_c(\ell)\,u_c(t).
\label{eq:zone-demand}
\end{equation}
We use a fixed serving map that assigns each demand-carrying zone $\ell$ to its nearest base station $b(\ell)$ and a dedicated subband $g(\ell)$, in place of the per-user association variables of~\cite{ZhouEtAlSPAtialTemporalEnergy2022}. Each base station has one subband for each zone it serves.

\subsection{Communications model and rate constraints}
\label{sec:comm-model}

Each base station operates over a licensed bandwidth of $\Wtot$ Hz, partitioned into subbands indexed by $g\in\setG$. Zones of the same base station are served simultaneously on their dedicated subbands, as in orthogonal frequency-division multiple access (OFDMA). For each time slot $t$, base station $b$, and subband $g$, the controller selects the allocated bandwidth $\Bv$ (Hz), the activation state $\xv\in\{0,1\}$, and the transmit power $\ptxv$ (W). The assignment $g(\ell)$ is fixed for the whole day, while the controller re-optimizes $\Bv$, $\xv$, and $\ptxv$ in every slot. Each subband is a contiguous group of physical resource blocks, consistent with the resource-block-group granularity in 3GPP TS~38.214~\cite{3gppTS38.214}. Turning a subband on or off models frequency-domain energy-saving adaptation as in 3GPP TR~38.864~\cite{3gppTR38.864}.

The total allocated bandwidth at each base station cannot exceed its licensed bandwidth, while inactive subbands are assigned zero bandwidth,
\begin{equation}
\textstyle\sum_{g} \Bv \le \Wtot,
\qquad \Bv \le \xv \Wtot,
\qquad \forall t,b,g.
\label{eq:bandwidth-sum}
\end{equation}
Likewise, with transmit-power budget $\PmaxBS$ per base station,
\begin{equation}
\textstyle\sum_{g} \ptxv \le \PmaxBS,
\quad 0 \le \ptxv \le \xv \PmaxBS,
\quad \forall t,b,g.
\label{eq:power-bs}
\end{equation}

We model the large-scale gain between base station $b$ and zone $\ell$ as~\cite{HeathLozanoFoundationsMIMOCommunications2018}
\begin{equation}
h_{b,\ell} =
\begin{cases}
h_0 \left(r_{b,\ell}/r_{0}\right)^{-\alpha}, & r_{b,\ell}\ge r_{0},\\
h_0, & r_{b,\ell} < r_{0},
\end{cases}
\label{eq:path-loss}
\end{equation}
where $h_0$ is the gain at the reference distance $r_0$, $r_{b,\ell}$ the distance between them, and $\alpha$ the path-loss exponent. The gain is a slot-average large-scale model without shadowing or fast fading.

We assume full frequency reuse, with subband $g$ occupying the same frequencies at every base station, so every other base station $j\neq b(\ell)$ transmitting on subband $g(\ell)$ interferes with zone $\ell$. With $N_0$ denoting the noise power spectral density (W/Hz), the signal-to-interference-plus-noise ratio (SINR) at zone $\ell$ is
\begin{equation}
\SINRtl =
\frac{\ptxvl\, h_{b(\ell),\ell}}
{N_0 \Bvl + \sum_{j \neq b(\ell)} \ptx_{t,j,g(\ell)}\, h_{j,\ell}}.
\label{eq:sinr-dedicated}
\end{equation}

We count the whole subband power of each interferer as interference, which is conservative when the interfering subband is wider than the served one. The resulting throughput of zone $\ell$ is $\Rtl = \Bvl \log_2(1 + \SINRtl)$. To satisfy the demand while limiting over-provisioning to a tolerance $\varepsilon$, we impose
\begin{equation}
\Dtl \;\le\; \Rtl \;\le\; (1+\varepsilon)\, \Dtl, \qquad \forall t,\ell.
\label{eq:qos-constraint}
\end{equation}

Let $\Psta$ denote the static power of a base station, $\Pact$ the power of each active subband, and $\beta$ the inverse of the power-amplifier efficiency. Following the linear power model of~\cite{AuerEtAlHowMuchEnergy2011}, the total power consumption of base station $b$ is
\begin{equation}
\Pcomm = \Psta + \Pact{\textstyle\sum_{g}} \xv + \beta{\textstyle\sum_{g}} \ptxv.
\label{eq:bs-power}
\end{equation}

\subsection{Energy model and carbon-adjusted electricity price}
\label{sec:energy-model}

Each site draws power from the grid or a local battery of capacity $\Emax$, which we treat as dedicated storage. Let $\pgridv$ denote the grid import, $\pchv$ the battery charging power, and $\pdisv$ the battery discharging power. At each time slot $t$, the power supply and consumption of each base station $b$ satisfy
\begin{equation}
\pgridv + \pdisv = \Pcomm + \pchv,
\quad \pgridv \ge 0,
\quad \forall t,b,
\label{eq:power-bal}
\end{equation}
where $\pgridv \ge 0$ prevents power export from the base station to the grid.

Let $\Ev$ denote the energy stored in the battery at base station $b$ during slot $t$, and let $\etac,\etad\in(0,1]$ denote the charging and discharging efficiencies. The battery energy evolves as
\begin{equation}
\label{eq:batt-energy}
\begin{aligned}
E_{t+1,b} &= \Ev+\etac \pchv\tau-\tfrac{1}{\etad} \pdisv\tau, \qquad \forall t,b,\\
E_{1,b} &= E_{T+1,b}=\Emax/2, \qquad \forall b.
\end{aligned}
\end{equation}
The stored energy and the battery powers satisfy
\begin{equation}
    \label{eq:batt-box}
    0 \le \Ev \le \Emax,\; 0 \le \pchv \le \Pchmax,\; 0 \le \pdisv \le \Pdismax,\;\; \forall t,b,
\end{equation}
so that the battery starts at half capacity and returns to it at the end of the day. To prevent simultaneous charging and discharging, we introduce $z_{t,b}\in \{0,1\}$ and impose 
\begin{equation}
\label{eq:batt-charging}
    \pchv \le \Pchmax z_{t,b}, \quad \pdisv \le \Pdismax(1-z_{t,b}), \quad \forall t,b.
\end{equation}

Let $\pi_t$ be the electricity price (USD/kWh), $\rho$ the carbon price (USD per metric ton of CO$_2$), and $\mu_t$ the grid carbon intensity (kg CO$_2$/kWh). Pricing the emissions behind each kilowatt-hour yields the carbon surcharge $\kappa_t=\rho\,\mu_t/1000$, where the factor of 1000 converts tons to kilograms. Adding the surcharge to the market price yields the carbon-adjusted electricity price as
\begin{equation}
\piEff=\pi_t+\kappa_t.
\label{eq:effective-price}
\end{equation}
The weight on emissions in $\piEff$ is thus the carbon price, a market quantity rather than a tuning parameter.

\section{Day-ahead planning and exact decomposition}
\label{sec:solve}

We assume perfect day-ahead forecasts of the demand $\Dtl$ and the grid signals $\pi_t$ and $\mu_t$, so the results bound from above what a forecast-driven scheduler achieves.

The controller minimizes the daily energy cost of base stations under the carbon-adjusted electricity price,
\begin{equation}
\min_{\substack{x,B,p^{tx},P^{grid}\\ p^{ch},p^{dis},E,z}} \;\; \sum_{t\in\setT}\sum_{b\in\setB}
\piEff\,\frac{\pgridv}{1000}\,\tau
\quad \text{s.t.}\quad
\eqref{eq:bandwidth-sum}, \eqref{eq:power-bs}, \eqref{eq:qos-constraint}\text{--}\eqref{eq:batt-charging},
\label{eq:original-formulation}
\end{equation}
where the factor of 1000 converts W to kW.

The resulting problem is mixed-integer and nonconvex. To avoid incentivizing unnecessary transmission when $\piEff<0$, the communication resources are chosen to minimize the site power in every slot whatever the sign of $\piEff$, while the battery may still charge at negative prices for later use. 

\emph{Proposition:} Under this rule, the optimization decomposes into two stages without loss of optimality. First, for each slot $t$, solve the resource subproblem
\begin{equation}
\min_{\actsym,B,\ptx}\; \textstyle\sum_b \Pcomm
\quad
\text{s.t.}\quad
\eqref{eq:bandwidth-sum}, \eqref{eq:power-bs}, \eqref{eq:qos-constraint}, \eqref{eq:bs-power},
\label{eq:resource-sub}
\end{equation}
and fix the resulting site powers $\Pcomm$. Then solve the day-level battery problem
\begin{equation}
\begin{aligned}
\min_{\pch,\pdis,E,z}\quad &
\textstyle\sum_{t,b}
\piEff(\pchv-\pdisv)\frac{\tau}{1000}
\quad
\text{s.t.}\quad
\eqref{eq:power-bal}\text{--}\eqref{eq:batt-charging}.
\end{aligned}
\label{eq:batt-lp}
\end{equation}

\begin{proof}
From the power balance \eqref{eq:power-bal}, $\pgridv=\Pcomm+\pchv-\pdisv$.
Hence, once $\Pcomm$ is fixed, its contribution to the objective is constant, and the remaining battery-dependent cost is exactly the objective in \eqref{eq:batt-lp}. Moreover, the constraint $\pgridv\ge0$ becomes
$\pdisv\le\Pcomm+\pchv$, which is included explicitly in the battery problem.
Since communication variables have no coupling across time slots, their minimum-power feasible allocation can be computed independently for each slot. When $\piEff\ge0$ in every slot, this plan is also optimal for~\eqref{eq:original-formulation}, since without export the battery can only offset the load of its own site and a higher site power cannot lower the cost. When $\piEff<0$ in some slots, the rule excludes the extra consumption that~\eqref{eq:original-formulation} would otherwise reward. With these site powers fixed, the only remaining temporal coupling is the battery dynamics, so \eqref{eq:batt-lp} gives the optimal battery schedule and the two-stage solution is exact under the stated rule, up to the per slot optimality gap of \slotGapMax{}.
\end{proof}

Each resource subproblem~\eqref{eq:resource-sub} is a small mixed-integer program with bilinear SINR constraints and a piecewise-linear logarithm, while~\eqref{eq:batt-lp} is a mixed-integer linear program.
Gurobi 12.0~\cite{gurobi} solves each resource subproblem in under a second to a gap of at most \slotGapMax{} and the battery problem once over the full day. The two-stage formulation thus solves a full day in about ten seconds, whereas solving \eqref{eq:original-formulation} directly yields no feasible solution within 40 minutes. We use one mobility realization throughout, on a 32-core desktop (Intel i9-14900KS) with a MIP gap of $10^{-4}$ and a 40-minute limit for \eqref{eq:original-formulation}.

\begin{figure}[!t]
  \centering
  \includegraphics[width=0.90\columnwidth]{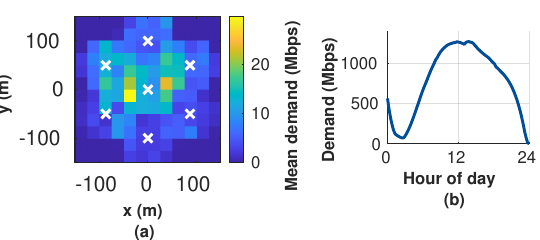}
  \caption{Spatiotemporal demand from the agent-based mobility pipeline. (a) Demand in each of the 144 zones of the $300\times300$~m map, averaged over the day, with the seven base stations marked by crosses. Only 42\% of the zone-slots carry nonzero demand. We retain the raw slot-to-slot variation without temporal smoothing. (b) Total demand over the day summed over all zones, which peaks near midday.}
  
  \label{fig:demand}
\end{figure}

\section{Numerical results}
\label{sec:experiments}

We evaluate the schedule on a seven-cell network with CAISO price and carbon-intensity data.
\begin{table}[!b]
\caption{Simulation parameters}
\label{tab:params}
\centering
    {\renewcommand{\arraystretch}{1.02}%
    \setlength{\tabcolsep}{4.5pt}%
    \footnotesize
        \begin{tabular}{c|c|c|c}
        \hline\hline
        Parameter & Value & Parameter & Value \\
        \hline
        $|\setB|$ & 7 (hex) & BS spacing & 100 m \\
        Map & $300\times300$ m & $L$ & 144 (25 m) \\
        $T$ & 96 & $\tau$ & 0.25 h \\
        $\Wtot$ & 100 MHz & $\PmaxBS$ & 200 W \\
        $h_0$ ($r_0{=}1$ m) & $-35$ dB & $\alpha$ & 2.5 \\
        $N_0$ & $3.2{\times}10^{-20}$ W/Hz & $\varepsilon$ & \qosMargin \\
        $\Psta$ & 40 W & $\Pact$ & 5 W \\
        $\beta$ & 2.8571 & $\etac=\etad$ & 0.9 \\
        $\Pchmax{=}\Pdismax$ & $\Emax/(2\,\mathrm{h})$ & Users & 600 (12:8:5:5) \\
        Peak rate (class C) & 1 Mbps/user & $\rho$ & \cpNow{} USD/ton \\
        \hline\hline
        \end{tabular}
    }
\end{table}

\subsection{Simulation setup}
\label{sec:exp-setup}
We use the seven-cell hexagonal layout of~\cite{ParkEtAlEnergyEfficiencyOptimization2026} with 144 zones of 25~m and the parameters of \tabref{tab:params}. The noise density corresponds to thermal noise with a 9~dB noise figure. Following \secref{sec:mobility-demand}, we draw users from the four daily activity templates of~\cite{ParkEtAlEnergyEfficiencyOptimization2026} (commuter, early commuter, shopper, and leisure, in the ratio 12:8:5:5) and sample their positions every 10~s. The APN class profiles peak near midday and fall to roughly 15\% of the peak overnight. We scale all classes by one factor so that the dominant class peaks at the per user rate of \tabref{tab:params}; the heavier class then peaks at 7.6~Mbps per user, and the network demand in \figref{fig:demand} peaks at 2.1~Mbps per user. The class profiles do not wrap at midnight, so the demand is close to zero in the last slots of the day.

\begin{figure}[!t]
  \centering
  \includegraphics[width=0.90\columnwidth]{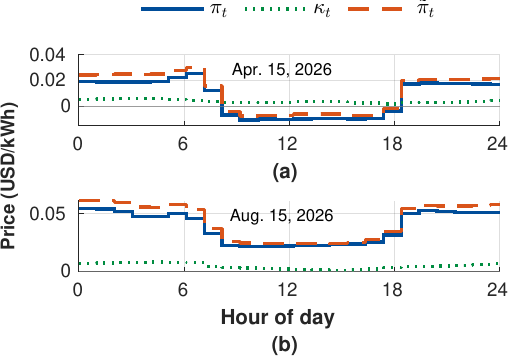}
  \caption{Electricity price $\pi_t$, carbon surcharge $\kappa_t$ at $\rho = 28.76$ USD/ton, and carbon-adjusted price $\tilde{\pi}_t$. (a) Spring reference day, Apr. 15, 2026, with negative prices during the midday solar surplus. (b) Summer reference day, Aug. 15, 2026, with positive prices throughout.}
  \label{fig:signals}
\end{figure}

The electricity price $\pi_t$ (USD/kWh) is the day-ahead locational marginal price at the \caisoNode{} trading hub of the California Independent System Operator (CAISO, OASIS PRC\_LMP report), while the grid carbon intensity $\mu_t$ (kg CO$_2$/kWh) is the system-average intensity from CAISO emission records~\cite{CAISOEmissions}. The carbon price is $\rho=\cpNow$~USD/ton, the 2025 California cap-and-trade allowance price~\cite{WorldBankStateTrends2025}.
As shown in \figref{fig:signals}(a), the \caisoDate{} price spans \priceRange{} and the carbon intensity \ciRange{}, with negative prices in the midday solar surplus. \figref{fig:signals}(b) shows the summer day \caisoDateAug{}, where prices remain positive.

\subsection{Emission reduction with battery storage}
\label{sec:exp-main}

\begin{figure}[t]
  \centering
  \includegraphics[width=0.90\columnwidth]{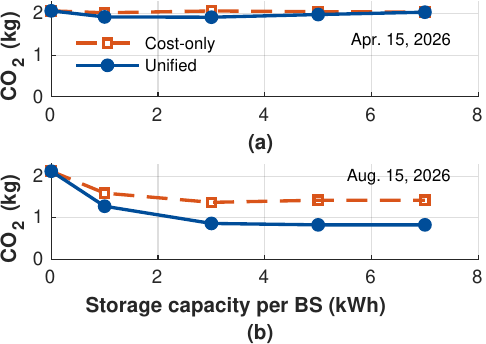}
  \caption{Daily network CO$_2$ emissions vs. storage capacity per BS $E_{\max}$, for the cost-only and unified objectives. (a) On the spring reference day, storage and carbon monetization have little effect on emissions. (b) On the summer reference day, emissions decrease with capacity, with larger reductions under the unified objective.}
  \label{fig:storage}
\end{figure}

We compare the unified objective \eqref{eq:original-formulation} with a cost-only baseline that solves the same problem with $\piEff$ replaced by the electricity price $\pi_t$, for storage capacities $\Emax\in\{0,1,3,5,7\}$~kWh per site. We do not compare with the association-based framework of~\cite{ZhouEtAlSPAtialTemporalEnergy2022}, since the fixed serving map removes the per-user association variables that framework optimizes. \figref{fig:storage} shows the resulting daily emissions on the two reference days. Without storage, the two objectives coincide. On the spring day, storage has little effect on emissions, and the unified objective reduces emissions by only \ablEmisCutNow{} relative to cost-only. This is because the midday low-price period already coincides with low carbon intensity, so carbon monetization adds little incentive beyond the electricity price alone.
On the summer day, emissions fall with storage and largely saturate by about 3~kWh per site, with the unified objective below cost-only at every capacity. At 3~kWh, the unified schedule reduces emissions by \ablEmisCutAug{} relative to cost-only and by \augBelowNoBattery{} relative to the battery-free case. Its electricity cost is \augElecPremium{} higher, while the total cost including the carbon surcharge is about 1\% lower. Here the low-price and low-carbon periods are less aligned, so the carbon surcharge shifts charging toward cleaner hours. All schedules serve the same traffic, so the differences come from when the grid energy is drawn.

\subsection{Sensitivity analysis of storage operation to carbon prices}
\label{sec:exp-cp}

\begin{figure}[t]
  \centering
  \includegraphics[width=0.90\columnwidth]{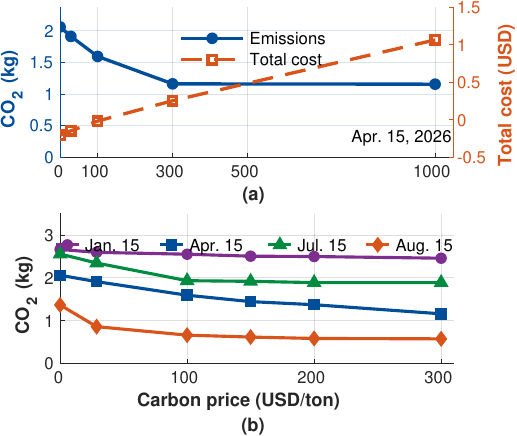}
  \caption{Daily network emissions vs. carbon price at $\Emax=3$~kWh per site. (a) Spring reference day, with the total cost on the right axis. (b) Seasonal repeats on four days. The largest reductions occur between \cpKneeLow{} and \cpKneeHigh~USD/ton.}
  \label{fig:cp}
\end{figure}

Emissions fall as the carbon price $\rho$ rises and saturate above \cpKneeHigh~USD per ton. \figref{fig:cp}(a) shows the effect at $\Emax=3$~kWh on the spring reference day, \caisoDate{}. The largest reduction occurs around $\cpKneeLow$--$\cpKneeHigh$~USD/ton, after which the curve saturates at about $\emisAtKnee$~kg. Below this range, the electricity price drives the battery, and above it, the carbon surcharge shifts charging toward lower-carbon hours. The total cost rises with $\rho$.

The same threshold appears on every day we test, while the size of the reduction varies with the daily price and carbon-intensity profiles. \figref{fig:cp}(b) repeats the analysis on Jan.~15, Jul.~15, and Aug.~15, 2026. The reduction ranges from \robWinterCut{} in winter to \robAugCut{} on the August day. Raising the carbon price to 1000~USD/ton lowers emissions by at most 1.1 percentage points more on any day, so the level reached by \cpKneeHigh~USD/ton is close to the attainable minimum at 3~kWh.

\subsection{Subband activation and battery storage operation}
\label{sec:behav}
\begin{figure}[!t]
  \centering
  \includegraphics[width=0.95\columnwidth]{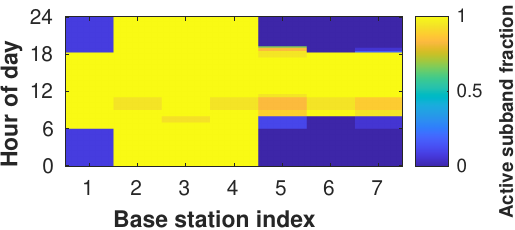}
  \caption{Active subband fraction across base station and time on the spring reference day under the unified objective at $E_{\max} = 3$ kWh. Activation follows the traffic demand.}
  \label{fig:heatmap}
\end{figure}

Subband activation follows the spatiotemporal demand. As shown in \figref{fig:heatmap}, more subbands are active at high-demand times (e.g., 6:00--18:00) and heavy-traffic base stations (e.g., index 2--4), while inactive zones switch off their assigned subbands. Under the fixed serving map, this activation pattern is determined by traffic demand rather than electricity price, so the emission reductions in Secs.~\ref{sec:exp-main} and \ref{sec:exp-cp} come from the battery schedule alone.

\begin{figure}[!t]
  \centering
  \includegraphics[width=0.95\columnwidth]{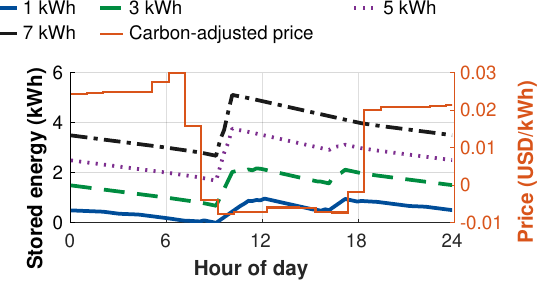}
  \caption{Stored energy averaged over the seven base stations on the spring reference day, under the unified objective at $\rho=\cpNow$~USD/ton for $\Emax=1$, 3, 5, and 7~kWh, with the carbon-adjusted price. Charging concentrates in the negative-price midday hours and discharge carries the evening peak.}
  \label{fig:soc}
\end{figure}

With communication resources determined by demand, the battery responds to the grid electricity and carbon intensity signals. \figref{fig:soc} shows the mean stored energy across base stations on the spring reference day for different storage capacities, together with the carbon-adjusted price at $\rho=\cpNow$~USD/ton. The batteries charge mainly during the low-price midday period and discharge during higher-price hours. Larger batteries shift more energy across time, but the benefit saturates near 3~kWh. Each site consumes only about 2.8~kWh per day and the battery cannot export to the grid, so a larger battery has no additional load to serve.

\section{Conclusion}
\label{sec:conclusion}
In this letter, we developed a carbon-aware framework for joint communication-resource and battery scheduling in cellular networks. A carbon-adjusted electricity price combines electricity cost and carbon emissions in a single objective, while the proposed decomposition separates slot-wise communication optimization from day-level battery scheduling without loss of optimality once transmission is not raised to exploit negative prices. Using real CAISO electricity-price and carbon-intensity traces, we showed that carbon-aware scheduling provides the largest emission reductions when low-price and low-carbon periods are misaligned, with greater benefits as storage capacity increases before saturating. We expect the framework to extend to massive MIMO base stations. Additional transceiver chains enter the power model in the same way as the subband activation power, as consumption that the resource subproblem switches off per slot. The two-stage decomposition is therefore unchanged. The site power, however, grows with the number of chains while the grid signals do not, so the battery capacity would have to scale with it to preserve the same relative emission reductions. The model omits battery degradation and capital cost. Future work can extend the framework to multi-region and full-year operation, incorporate electricity-price and carbon-intensity forecasts through rolling-horizon re-planning, and allow flexible user--base station association.

\bibliographystyle{IEEEtran}
\bibliography{JP_carbon_refs}

\end{document}